\documentclass[%
 reprint,
superscriptaddress,
nofootinbib,
 amsmath,amssymb,
 aps,
prd,
floatfix,
]{revtex4-1}

\usepackage{graphicx}
\usepackage{dcolumn}
\usepackage{hyperref}
\usepackage{multirow}

\usepackage{xcolor}
\definecolor{linkcolor}{rgb}{0.0, 0.47, 0.75}
\definecolor{citecolor}{rgb}{1.0, 0.5, 0.0}
\hypersetup{
  linkcolor  = linkcolor,
  citecolor  = linkcolor,
  urlcolor   = linkcolor,
  colorlinks = true
}

\begin{document}
\title{Quantifying the tension between cosmological and terrestrial constraints on neutrino masses}

\author{Stefano Gariazzo}
\affiliation{Istituto Nazionale di Fisica Nucleare (INFN), Sezione di Torino, Via P.\ Giuria 1, I-10125 Turin, Italy}

\author{Olga Mena}%
\affiliation{Instituto de F{\'\i}sica Corpuscular  (CSIC-Universitat de Val{\`e}ncia), E-46980 Paterna, Spain}

\author{Thomas Schwetz}
\affiliation{Institut f\"ur Astroteilchenphysik, Karlsruher Institut f\"ur Technologie (KIT), Hermann-von-Helmholtz-Platz 1, 76344 Eggenstein-Leopoldshafen, Germany}


\newcommand{\mnu}{\ensuremath{\sum m_\nu}}
\newcommand{\dmsol}{\ensuremath{\Delta m^2_{21}}}
\newcommand{\dmatm}{\ensuremath{|\Delta m^2_{31}|}}

\begin{abstract}
The sensitivity of cosmology to the total neutrino mass scale $\mnu$ is approaching the minimal values required by oscillation data. 
We study quantitatively possible tensions between current and forecasted cosmological and terrestrial neutrino mass limits by applying suitable statistical tests such as Bayesian suspiciousness, parameter goodness-of-fit tests, or a parameter difference test. In particular, the tension will depend on whether the normal or the inverted neutrino mass ordering is assumed. We argue, that it makes sense to reject inverted ordering from the cosmology/oscillation comparison only if data are consistent with normal ordering. Our results indicate that, in order to reject inverted ordering with this argument, an accuracy on the sum of neutrino masses $\sigma ({m_\nu})$ of better than 0.02~eV would be required from future cosmological observations. 
\end{abstract}

\maketitle

\section{Introduction}

Massive neutrinos affect cosmological observables due to the unique behaviour as dark radiation in early times and as dark matter at late times, see \cite{Lesgourgues:2006nd} for a review.
Effectively, cosmology is sensitive to the total energy density of relic neutrinos.
In the standard scenario, after they become non-relativistic,
the correspondence between non-relativistic neutrino energy density and neutrino masses is approximately given by~\cite{Froustey:2021qqq}
\begin{equation}
\Omega_\nu h^2 = \sum_i m_i/(93.12\text{ eV})\,,
\end{equation}
where $h$ is the reduced Hubble parameter.
Constraining the non-relativistic neutrino energy density, therefore,
allows us to obtain bounds on the sum of the neutrino masses
under the assumption that the above equation holds. Combining information from CMB and BAO observations, the Planck collaboration obtains~\cite{Aghanim:2018eyx}
\begin{equation}\label{eq:Planck}
  \sum m_\nu \equiv \sum_{i=1}^3m_i < 0.12\,{\rm eV} \,(95\%\,{\rm CL}) \,.
\end{equation}
Adding more recent data, even more stringent limits can be obtained. For instance, Ref.~\cite{DiValentino:2021hoh} finds
\begin{equation}\label{eq:cosmo_bound}
  \sum m_\nu < 0.09\,{\rm eV} \,(95\%\,{\rm CL}) \, ,
\end{equation}
see also \cite{Palanque-Delabrouille:2019iyz,diValentino:2022njd}. In the near future we expect that the 
DESI~\cite{DESI:2016fyo} and/or Euclid~\cite{Amendola:2016saw} surveys  may provide sensitivities to $\sum m_\nu$ of down to $0.02\,\mathrm{eV}$ or beyond, see e.g.
\cite{Font-Ribera:2013rwa,Basse:2013zua,Hamann:2012fe,Carbone:2010ik,Brinckmann:2018owf}. 

On the other hand, neutrino oscillation experiments provide accurate determinations of the two neutrino mass-squared splittings \cite{deSalas:2020pgw}, see also Refs.~\cite{Esteban:2020cvm,Capozzi:2021fjo}:
\begin{equation}\label{eq:oscillations}
\begin{split}
\Delta m^2_{21} &=(7.50\pm 0.21)\times 10^{-5} \,{\rm eV}^2 \,, \\
|\Delta m^2_{31}|&= \left\{
\begin{array}{ll}
(2.550\pm0.025)\times 10^{-3}\, {\rm eV}^2 &\quad \rm (NO) \\
(2.450\pm0.025)\times 10^{-3}\, {\rm eV}^2 &\quad \rm (IO)
\end{array}
\right. \,,
\end{split}    
\end{equation}
where $\Delta m^2_{ij} \equiv m_i^2-m_j^2$. The sign of $\Delta m^2_{31}$ determines the type of neutrino mass ordering,  being positive for normal ordering (NO) and negative for inverted ordering (IO).
With the mass-splittings determined, oscillation data provide a lower bound on the sum of the neutrino masses, obtained by assuming that the lightest neutrino mass is zero. From Eq.~\eqref{eq:oscillations}, one finds
%
\begin{equation}\label{eq:lower_bound}
    \begin{split}
        \sum m_\nu > \left\{
        \begin{array}{ll}
(0.0591 \pm 0.00027) \, {\rm eV} &\quad \rm (NO) \\
(0.0997 \pm 0.00051) \, {\rm eV} &\quad \rm (IO) \\
        \end{array}\right.
    \end{split}\,.
\end{equation}
Comparing these results with Eqs.~\eqref{eq:Planck} and \eqref{eq:cosmo_bound}, it is possible to notice that cosmological upper bounds are already comparable to the lower bound for IO, and near future sensitivities will probe the NO region case as well.
This may happen in two ways: by measuring a value equal or slightly larger than $\mnu\simeq0.06$~eV and confirming that at least two neutrinos have a positive mass in agreement with oscillation results, or by strengthening the current upper bounds to the level that both minimal values in Eq.~\eqref{eq:lower_bound} will be disfavored by cosmology.
Therefore, it is mandatory  to quantify a possible tension between cosmology and oscillation data, which constitutes the main goal of this manuscript. Such a tension would have important implications: the absence of a detection of a finite neutrino mass in cosmology as predicted by Eq.~\eqref{eq:lower_bound} could be a striking signal for a non-standard cosmological model beyond the vanilla $\Lambda$CDM model and/or non-standard neutrino properties, see e.g.~\cite{Alvey:2021xmq} for a discussion. 

Furthermore, the tension between the lower bound on $\sum m_\nu$ for IO and the bound from cosmology could be used in principle to disfavour IO compared to NO, see e.g.  \cite{Hannestad:2016fog,Gerbino:2016ehw,Vagnozzi:2017ovm,Simpson:2017qvj,Gariazzo:2018meg,
DeSalas:2018rby,Heavens:2018adv,Gariazzo:2018pei,RoyChoudhury:2019hls,Mahony:2019fyb,
Hergt:2021qlh,Jimenez:2022dkn,Gariazzo:2022ahe} for an incomplete list of studies on this topic.
Typically, in these papers some kind of Bayesian model comparison between NO and IO is performed, leading to posterior odds for the two models. 

In the following we will address this question with a slightly different approach, namely by quantifying the tension between cosmology and oscillation data for the two orderings. We argue that it is meaningful to reject IO from a comparison of cosmology and oscillations \emph{only if} these two data sets are consistent for NO. In the case when there is tension between cosmology and oscillations for both orderings, a relative comparison of the two models can be misleading. We shall explore this putative tension exploiting both current and future cosmological measurements.

The structure of the manuscript is as follows. In Sec.~\ref{sec:metrics} we describe the different methods commonly exploited in the literature to quantify tensions between two sets of measurements. Section \ref{sec:analysis} contains a description of the methodology for the numerical analyses, the parameterizations employed to describe the parameter space and the data involved in quantifying the tension between cosmological and terrestrial neutrino mass measurements. Section~\ref{sec:results} presents the results from our analyses, including a mass ordering comparison. Finally, we conclude in Sec.~\ref{sec:conclusions}.

\section{Tension metrics}\label{sec:metrics}

In this section we provide a brief review of various metrics used to quantify a tension between different data sets. We follow closely the discussion in Ref.~\cite{DES:2020hen}, where a number of tests is reviewed and applied in the context of the $H_0$ tension. We refer the interested reader to Ref.~\cite{DES:2020hen} for further references and more in depth discussions of the various tests. Additional discussions can be found for instance, in the context of cosmology in \cite{Lin:2019zdn,Raveri:2018wln}, in the context of Type Ia Supernova analysis in \cite{Amendola:2012wc}, and within a frequentist framework in the context of neutrino oscillations in \cite{Maltoni:2003cu}.

To fix the notation, in the following
$\mathcal{L}_D = P(D|\theta,M)$ denotes the likelihood, which is the probability for the data $D$ given a model $M$ with parameters $\theta$, $\Pi = P(\theta|M)$ is the prior for the parameters,
\begin{equation}
\mathcal{Z}_D = P(D|M) = \int \text{d}\theta\, \mathcal{L}_D(\theta) \, \Pi(\theta)~,    
\end{equation}
is the Bayesian evidence, and  
\begin{equation}
\mathcal{P}_D(\theta) = P(\theta | D, M) = \frac{\mathcal{L}_D(\theta) \, \Pi(\theta)}{\mathcal{Z}_D}~,   
\end{equation}
is the posterior density for the parameters $\theta$ for data $D$. Considering now two data sets $D=A,B$, the question posed here is whether these data are consistent within a given model. In order to quantitatively address this question, the following tests can be used:
\begin{itemize}
\item \textbf{Bayesian evidence ratio.} Consider the ratio
\begin{equation}
\label{eq:lnZratio}
R\equiv
\frac{\mathcal{Z}_{AB}}{\mathcal{Z}_{A}\mathcal{Z}_{B}} \,.
\end{equation}
The numerator corresponds to the evidence when data sets $A$ and $B$ are described by the same set of parameters $\theta$, whereas in the denominator different parameters may be preferred by the two data sets. Values of $R \gg 1 \, (\ll 1)$ would indicate agreement (disagreement) between the two data sets. As discussed in \cite{DES:2020hen}, $R$ is dependent on the prior volume, and small values of $R$,  indicating a possible tension between data sets, can be increased by increasing the prior volume. Therefore, we will not use the Bayesian evidence ratio in our tension analysis below.

\item \textbf{Bayesian suspiciousness.} 
This test departs from the Bayesian evidence ratio, but the information ratio $I$ based on the Kullback-Leibler divergence is used to remove the prior dependence.
Consider the log-information ratio
\begin{equation}
\label{eq:lnI}
\ln I = \mathcal{D}_A+\mathcal{D}_B-\mathcal{D}_{AB}\,,
\end{equation}
where the Kullback-Leibler divergence is defined as
\begin{equation}
\label{eq:KL_div}
\mathcal{D}_D
=
\int \text{d}\theta \, 
\mathcal{P}_D
\ln\left(\frac{\mathcal{P}_D}{\Pi}\right) \,.
\end{equation}
Using the log-information ratio 
we can cancel the prior dependence from the Bayesian evidence ratio $R$
and define the suspiciousness \cite{Handley:2019wlz}:
\begin{equation}
\label{eq:lnS}
\ln S
\equiv
\ln R - \ln I \,.
\end{equation}
As for $R$, positive values of $\ln S$ indicate agreement among the data sets while negative ones indicate disagreement.

For Gaussian posteriors, the quantity $(d-2\ln S)$ follows a $\chi_d^2$ distribution,
where the number of degrees of freedom can be obtained 
using the Bayesian model dimensionality defined in \cite{Handley:2019pqx}:
\begin{equation}
\label{eq:bmd}
d_D = 2
\int
\text{d}\theta \,
\mathcal{P}_D
\left[
\ln\left(\frac{\mathcal{P}_D}{\Pi}\right)
-\mathcal{D}_D
\right]^2 \,.
\end{equation}
In order to compute the significance of the tension between two data sets, the relevant Bayesian dimensionality can be obtained using \cite{Handley:2019pqx}
\begin{equation}\label{eq:d}
  d=d_A+d_B-d_{AB} \,.   
\end{equation}

As we will discuss in the following, the Bayesian model dimensionality $d_D$
may have problems when dealing with posteriors that are not
Gaussian in the parameters under consideration,
or when the prior limits impose a significant cut on the posterior shape.
In these cases,
we will replace the Bayesian model dimensionality with a more naive counting for the number of degrees of freedom, see below.

\item \textbf{Parameter goodness-of-fit tests.} This test is based on the idea to evaluate the ``cost'' of explaining data sets together (i.e., with the same parameter values) as compared to describing them separately (i.e., each data set can chose its own preferred parameter values). Therefore this type of tests is sometimes also called ``goodness-of-fit loss'' tests.
We take as an example two data sets $A$ and $B$, as of interest in this study (generalization to more data sets is straight-forward). Compatibility of the data sets is evaluated using the test statistic
\begin{equation}\label{eq:PG}
\begin{split}
    Q \equiv &-2\ln\mathcal{L}_{AB}(\hat\theta_{AB}) \\
    &+ 2\ln\mathcal{L}_{A}(\hat\theta_{A}) + 2\ln\mathcal{L}_{B}(\hat\theta_{B}) \,.
\end{split}
\end{equation}
Here $\hat\theta_D$ denotes the parameter values which ``best'' describe data set $D$. 
In the context of frequentist statistics, $\hat\theta_D$ is taken as the parameter values maximizing the likelihood $\mathcal{L}_D(\theta)$~\cite{Maltoni:2003cu}. In this case, the test statistic is denoted as $Q\equiv\chi^2_{\rm PG}$ and, by construction, $\chi^2_{\rm PG}$ is independent of the prior and any re-parameterization (as long as the number of independent parameters remains the same).  In the context of Bayesian analysis, $\hat\theta_D$ is taken at the parameter values at the ``maximum a posteriori'' (MAP, the point at which the posterior assumes its maximum value), which in general does depend on the prior choice, see Refs.~\cite{DES:2020hen,Raveri:2018wln}, where the corresponding test statistic is denoted by $Q_{\rm DMAP}$ (difference of log-likelihoods at their MAP point). For flat uninformative priors ($\Pi(\theta) = const$) maximum likelihood and maximum posterior points are identical and $\chi^2_{\rm PG} = Q_{\rm DMAP}$.

Under certain regularity conditions, $Q$ from Eq.~\eqref{eq:PG} is distributed as a $\chi^2_n$ distribution, where $n$ is the number of parameters in common to both data sets $A$ and $B$, 
\begin{equation}\label{eq:n}
  n=p_A+p_B-p_{AB} \,,   
\end{equation}
where $p_D$ denotes the parameters of data set $D$, see \cite{Maltoni:2003cu,Raveri:2018wln} for precise definitions. 

\item \textbf{Parameter differences.}
This test measures the distance between posterior distributions for the parameters $\theta$,
given two different datasets \cite{Raveri:2021wfz,Raveri:2019gdp}.
Let us define the difference $\Delta\theta=\theta_1 - \theta_2$,
where $\theta_1$ and $\theta_2$ are two points in the shared parameter space.
Assuming, as in our case, that the datasets $A$ and $B$ are independent,
the posterior distribution for $\Delta\theta$ can be computed using:
\begin{equation}
\mathcal{P}_\Delta(\Delta\theta)
=
\int
\mathcal{P}_A(\theta)\mathcal{P}_B(\theta-\Delta\theta)
\,
d\theta
\,,
\end{equation}
where the integral runs over the entire parameter space of the shared parameters.
The probability that there is a parameter shift between the two posteriors
is quantified by the posterior mass above the iso-contour of no shift ($\Delta\theta=0$).
This can be obtained by performing the following integral:
\begin{equation}
\label{eq:delta_par_shift}
\Delta
=
\int_{\mathcal{P}_\Delta(\Delta\theta)>\mathcal{P}_\Delta(0)}
\mathcal{P}_\Delta(\Delta\theta)
\,
d\Delta\theta
\,,
\end{equation}
which is symmetric for changes of datasets A$\leftrightarrow$B and gives us the probability $\Delta$.
If $\Delta$ is close to zero, no shift is present and the two datasets are in agreement.
On the contrary, a probability $\Delta$ close to one indicates a tension between datasets.
\end{itemize}

For all the tests considered below we will report 
significance in terms of number of standard deviations
by converting probabilities into two-sided Gaussian standard deviations.

\section{Analysis}
\label{sec:analysis}

\subsection{Technical details}
One of the objectives of this analysis is to compute the Suspiciousness tests
for which the calculation of Bayesian evidences and Kullback-Leibler divergences is required.
In order to obtain such quantities, we perform our numerical scans with
\texttt{PolyChord}~\cite{Handley:2015aa}
and analyse the results using \texttt{anesthetic}~\cite{Handley:2019mfs}.
Implementations of other tests are taken from the code \texttt{Tensiometer}~\footnote{\url{https://github.com/mraveri/tensiometer}.}.
Concerning the implementation of the parameter differences test,
we adopt the \texttt{Tensiometer} when considering multi-dimensional parameter spaces,
while we directly implement the integral in Eq.~\eqref{eq:delta_par_shift}
when dealing with only one parameter.

Our numerical implementation considers 
different parameterizations (see later) for the neutrino masses,
which are constrained using a set of cosmological and terrestrial observations.
In order to reduce the random fluctuations that arise from the initial sampling of the live points
in \texttt{PolyChord},
we repeat the nested sampling runs several times for each data combination and 
neutrino mass parameterization,
varying the number of live points each time between 500 and 1500.
The quoted results are taken as the mean of the tension metrics applied to each run separately.

\subsection{Parameterizations}
\label{subs:parameterizations}

Our interest below is focused on studying different constraints on neutrino masses.
Considering a model with three massive neutrinos, there are several possible ways
to describe their mass spectrum that have been adopted in the past in the context of cosmological studies, e.g., 
\cite{Hannestad:2016fog,Simpson:2017qvj,Schwetz:2017fey,Gariazzo:2018pei,Heavens:2018adv,Jimenez:2022dkn,Gariazzo:2022ahe}. Below we will present results for two representative examples, denoted by ``$3\nu$'' and 
``$\Sigma$'', respectively:
\begin{itemize}
\item $\mathbf{3\nu}$\textbf{-parameterization:} 
We consider the three neutrino masses $m_A$, $m_B$, $m_C$
as independent parameters in the analyses.
After sampling the parameters, the masses are ordered from the smallest to the largest and
assigned to the mass eigenstates, depending on the considered mass ordering:
$m_1<m_2<m_3$ for NO, $m_3<m_1<m_2$ for IO.
Similar to \cite{Jimenez:2022dkn} (see also \cite{Schwetz:2017fey,Gariazzo:2022ahe}),
we impose a Gaussian prior on the logarithm of the three neutrino masses,
with the same mean $\mu$ and standard deviation $\sigma$. Hence, neutrino masses are sampled according to a log-normal distribution, without any prior boundaries.\footnote{We have tested alternative sampling methods, such as sampling the masses or the logarithms of the masses uniformly within a given range and then apply the lognormal distribution \cite{Gariazzo:2022ahe}, leading to similar results.} The mean $\mu$ and standard deviation $\sigma$ are hyper-parameters in the analysis. We sample them considering a uniform prior on their logarithm, with bounds $5\cdot10^{-4}<\mu/\text{eV}<0.3$ and $5\cdot10^{-3}<\sigma/\text{eV}<20$, respectively, and marginalize over them.

\item $\mathbf{\Sigma}$\textbf{-parameterization:} We describe the neutrino masses by means of their sum \mnu\ and the two
mass splittings \dmsol\ and \dmatm~\cite{Heavens:2018adv}.
As, for practical purposes, the current and future cosmological probes considered here only depend on \mnu, it is possible to marginalize first the likelihood of terrestrial data over the two mass splittings and then perform the combined analysis or the compatibility analysis with just one free parameter (\mnu).
We verified that this procedure leads to very similar results as performing the entire calculation with three free parameters (\mnu, \dmsol, \dmatm).
For definiteness we show here only the results sampling \mnu\ with a linear prior, since our checks using a logarithmic prior provide very similar results.
\end{itemize}

Following \cite{Gariazzo:2018pei}, we have considered also a range of other parameterizations, e.g., using either \mnu\ or the lightest neutrino mass and the two mass splittings \dmsol\ and \dmatm; with uniform prior distribution either on the parameters themselves or on their logarithm.
We identified our benchmark parameterizations $3\nu$ and $\Sigma$ described above as representative examples, and therefore we restrict the discussion to the two of them.

\subsection{Cosmological and terrestrial information on neutrino masses}
\label{subs:data}

The aim of this study is to determine the level of tension between cosmological measurements
of neutrino masses
and terrestrial constraints on the masses and mass splittings.
For that purpose, we shall consider the following data constraints:
\begin{itemize}
\item Neutrino oscillation constraints are simulated by a Gaussian likelihood
on the solar and atmospheric mass differences with mean and standard deviations according to Eq.~\eqref{eq:oscillations}. Note that the $\Delta\chi^2$
between NO and IO from oscillation data does not affect the tension metrics and is therefore not relevant for our analyses.
\item For terrestrial neutrino mass measurements we include the result from
KATRIN by adopting a Gaussian likelihood with \cite{KATRIN:2021uub}
\begin{equation}
\label{eq:mbeta}
m_\beta^2 = (0.06 \pm 0.32) \, \text{eV}^2 \,.
\end{equation}
The region of interest corresponds to quasi-degenerate neutrinos and we can use the approximation for the effective mass parameter relevant for KATRIN:
\begin{equation}
m_\beta^2 \approx \left(\frac{\sum m_\nu}{3}\right)^2 \,.
\end{equation}
Effectively, this provides an upper bound on $\sum m_\nu$ for the terrestrial data.
\end{itemize}
The combination of these two data sets is denoted as ``terrestrial'' in the following. For the cosmological data we consider current data, as well as two possible future scenarios:
\begin{itemize}
\item For current cosmological data, we consider the full posterior distribution obtained using
Planck temperature, polarization and lensing data together with Supernovae Ia luminosity distance measurements and Baryon Acoustic Oscillations plus Redshift Distortions from SDSS IV, which corresponds to a 95\% CL upper limit
$\mnu<0.09$~eV~\cite{DiValentino:2021hoh}.
\item For future cosmological probes, we shall consider a precision of 0.02~eV on the sum of neutrino masses \cite{Font-Ribera:2013rwa,Basse:2013zua,Hamann:2012fe,Carbone:2010ik,Brinckmann:2018owf}  and two alternative scenarios: either a value for \mnu\ corresponding to the minimal value as predicted for the NO, see Eq.~\eqref{eq:lower_bound},
\begin{equation}\label{eq:futureNO}
    \mnu=0.06\pm0.02\, \text{eV \quad (future NO)}\,,
\end{equation}
or a hypothetical non-observation of finite neutrino masses in cosmology, 
\begin{equation}\label{eq:future0}
    \mnu=0 \pm0.02\, \text{eV \quad (future 0)}\,.
\end{equation}
Note that the latter case, by construction, is in tension with oscillation data. We will use the statistical tests discussed above to quantify this statement. In both cases, we assume a Gaussian likelihood for \mnu.
\end{itemize}

\begin{figure}
\centering
\includegraphics[width=\columnwidth]{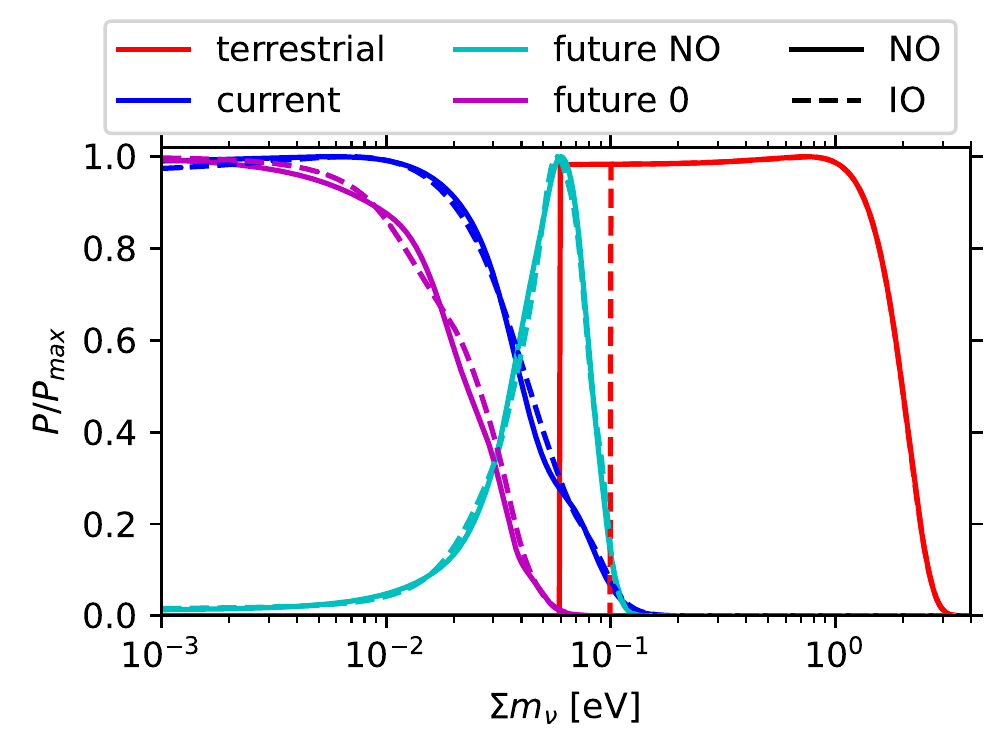}
\caption{Posterior of $\mnu$ for NO (IO) in solid (dashed) lines,
given different data sets, either terrestrial or cosmological ones (no combinations are shown).
}
\label{fig:S_mnu_post}
\end{figure}

Figure~\ref{fig:S_mnu_post} shows the posteriors for various data sets using the $\Sigma$-parameterization. We observe the top-hat shaped distribution for terrestrial data (red curves), with the lower bound provided by oscillations (its value depending on NO or IO) and the upper bound provided by KATRIN. The interplay with the assumed cosmological data sets is apparent from the figure, and below we are going to quantify possible existing tensions among them.

\section{Results}
\label{sec:results}

\begin{table*}[t]
\centering
\begin{tabular}{l|l||c|c||c|c||c||c|c|c|c}
Cosmo data & Model & $\ln S$ & $p_S$ & $Q_{DMAP}$ & $1-P(Q_{DMAP})$ & $1-P(\Delta)$ & $d_A$ & $d_B$ & $d_{AB}$ & $d$ \\ \hline
\hline\multirow{4}*{current} & $3\nu$ NO & -1.34 & 0.13 ($1.52\sigma$) & 2.62 & 0.45 ($0.75\sigma$) & 0.011 ($2.66\sigma$) & 1.94 & 0.31 & 2.28 & -0.02 \\
 & $3\nu$ IO & -3.30 & 0.023 ($2.29\sigma$) & 5.61 & 0.13 ($1.50\sigma$) & 0.0014 ($3.90\sigma$) & 1.98 & 0.33 & 3.54 & -1.23 \\
\cline{2-11}
 & $\Sigma$ NO & -1.23 & 0.063 ($1.86\sigma$) & 2.62 & 0.098 ($1.65\sigma$) & 0.021 ($2.31\sigma$) & 0.44 & 1.24 & 1.47 & 0.21 \\
 & $\Sigma$ IO & -2.86 & 0.0095 ($2.59\sigma$) & 5.59 & 0.017 ($2.39\sigma$) & 0.0014 ($3.20\sigma$) & 0.46 & 1.21 & 1.65 & 0.02 \\
\hline
\hline\multirow{4}*{future NO} & $3\nu$ NO & 0.55 & 0.59 ($0.53\sigma$) & 0.033 & 1 ($0.00\sigma$) & 0.23 ($1.19\sigma$) & 1.94 & 1.81 & 2.32 & 1.43 \\
 & $3\nu$ IO & -1.75 & 0.089 ($1.70\sigma$) & 3.99 & 0.26 ($1.12\sigma$) & 0.018 ($2.54\sigma$) & 1.98 & 1.71 & 2.76 & 0.93 \\
\cline{2-11}
 & $\Sigma$ NO & 0.2 & 0.44 ($0.78\sigma$) & 0.035 & 0.84 ($0.21\sigma$) & 0.1 ($1.62\sigma$) & 0.44 & 0.93 & 1.00 & 0.38 \\
 & $\Sigma$ IO & -2.16 & 0.021 ($2.31\sigma$) & 3.98 & 0.043 ($2.03\sigma$) & 0.00099 ($3.29\sigma$) & 0.46 & 0.92 & 1.60 & -0.22 \\
\hline
\hline\multirow{4}*{future 0} & $3\nu$ NO & -4.58 & 0.0068 ($2.70\sigma$) & 8.78 & 0.032 ($2.14\sigma$) & 0.0016 ($3.66\sigma$) & 1.94 & 0.31 & 2.55 & -0.30 \\
 & $3\nu$ IO & -13.04 & 2.3e-06 ($4.74\sigma$) & 24.90 & 1.6e-05 ($4.31\sigma$) & 2.5e-05 ($5.28\sigma$) & 1.98 & 0.32 & 3.51 & -1.21 \\
\cline{2-11}
 & $\Sigma$ NO & -4.56 & 0.0015 ($3.18\sigma$) & 8.80 & 0.0028 ($2.99\sigma$) & 8.1e-06 ($4.46\sigma$) & 0.44 & 1.00 & 1.71 & -0.26 \\
 & $\Sigma$ IO & -12.68 & 2.8e-07 ($5.13\sigma$) & 24.91 & 5.4e-07 ($5.01\sigma$) & 4.1e-10 ($6.25\sigma$) & 0.46 & 1.03 & 1.87 & -0.37 \\
\hline
\end{tabular}
\caption{Tension between cosmological and terrestrial neutrino mass determination assuming different cosmological data sets: current data, a future observation with a value for \mnu\ consistent with NO (future NO, Eq.~\eqref{eq:futureNO}), and a non-observation of \mnu\ (future 0, Eq.~\eqref{eq:future0}). In each case we show results for the two parameterizations $3\nu$ and $\Sigma$ (see section~\ref{subs:parameterizations}) and the two mass orderings. The table shows the test statistics $\ln S$, $Q_{\rm DMAP}$ from Eqs.~\eqref{eq:lnS} and \eqref{eq:PG} respectively, and the probabilities of the data sets being consistent, $p_S$, $1-P(Q_{\rm DMAP})$ and  $1-P(\Delta)$, corresponding to the suspiciousness test, the parameter goodness-of-fit test, and the parameter shift test, respectively. In the right part of the table we show the Bayesian model dimensionalities according to Eqs.~\eqref{eq:bmd} and \eqref{eq:d}, indicating with $A$ the terrestrial and $B$ the cosmological data sets. The values for $p_S$ [as well as for $1-P(Q_{\rm DMAP})$] are calculated with the parameter counting according to Eq.~\eqref{eq:n}, i.e., for 3 (1) dof for the $3\nu$ ($\Sigma$) parameterization.}
\label{tab:summary}
\end{table*}

Let us now present the results of our analysis about the consistency or possible tension between cosmology and terrestrial neutrino mass determinations. The numerical results for the suspiciousness, parameter goodness-of-fit, and parameter shift tests are summarized in Tab.~\ref{tab:summary}. We show the results for the corresponding test statistics as well as significances. The compatibility is tested assuming either the current cosmological likelihood, or possible future determinations of \mnu, with the two cases future~NO and future~0 discussed in section~\ref{subs:data}. Furthermore, we check how the results depend on the type of the neutrino mass ordering (normal versus inverted) as well as on the parameterization used for the neutrino masses ($3\nu$ versus $\Sigma$, see section~\ref{subs:parameterizations}).

\subsection{Suspiciousness and parameter goodness-of-fit tests}

\begin{figure*}
\centering
\includegraphics[width=0.8\columnwidth]{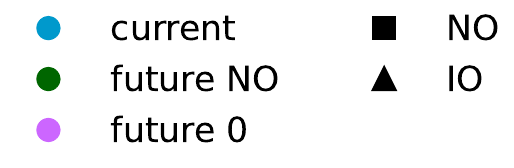}\\
\includegraphics[width=0.99\columnwidth]{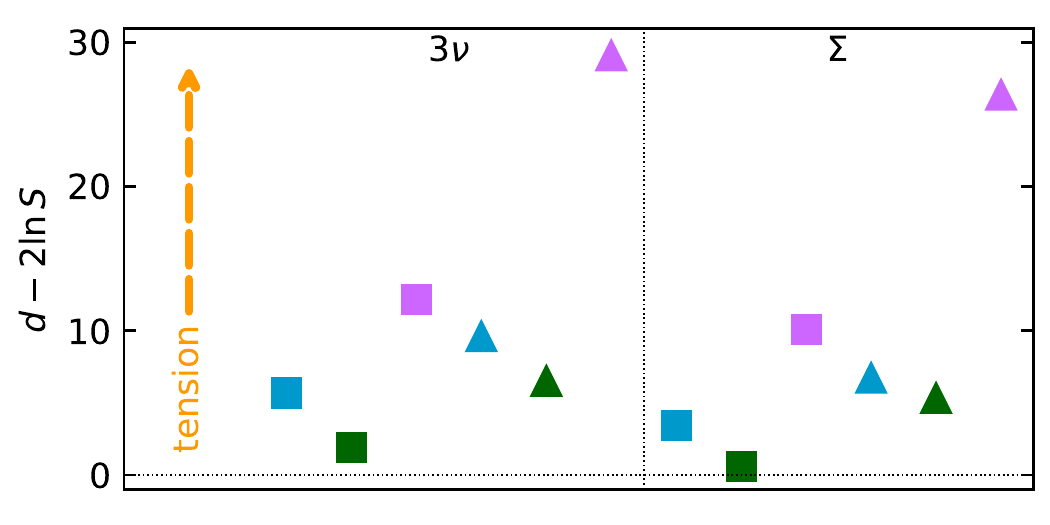}$\qquad$
\includegraphics[width=0.99\columnwidth]{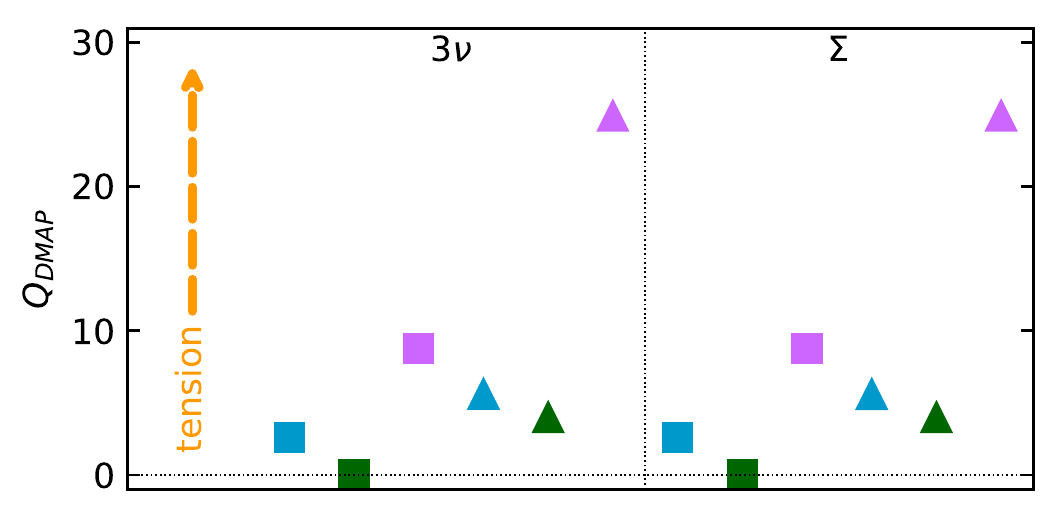} \\
\includegraphics[width=0.99\columnwidth]{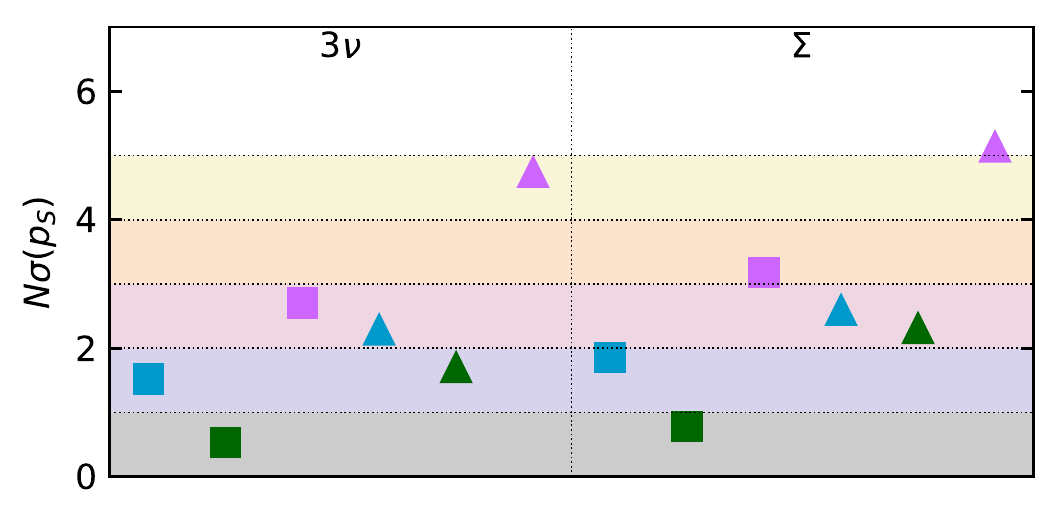}$\qquad$
\includegraphics[width=0.99\columnwidth]{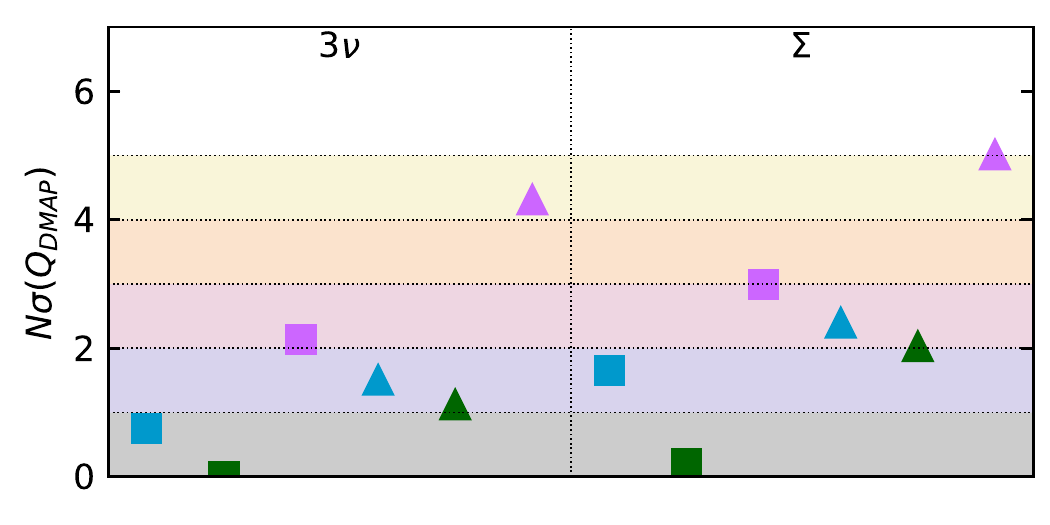}
\caption{Tension between cosmological and terrestrial experiments according to the suspiciousness test  (left panels) and parameter goodness-of-fit test (right panels). Upper panels show the corresponding test statistics as defined in Eqs.~\eqref{eq:lnS} and \eqref{eq:PG}. The left (right) part in each panel corresponds to the $3\nu$ ($\Sigma$) parameterization as defined in Sec.~\ref{subs:parameterizations}. 
Lower panels show the corresponding significance in numbers of standard deviations obtained by converting to a $p$-value assuming a $\chi^2_d$ distribution, where $d = 3$ (1) for the 
$3\nu$ ($\Sigma$) parameterization. Different colors indicate different assumptions on the  the cosmological data set, see Sec.~\ref{subs:data}, and square (triangle) symbols correspond to normal (inverted) neutrino mass ordering.}
\label{fig:lnS-PG}
\end{figure*}

We start by discussing the results for the suspiciousness and the parameter goodness-of-fit test, which are illustrated graphically in Fig.~\ref{fig:lnS-PG}. In the upper panels we show the corresponding test statistics $(d-2\ln S)$ and $Q_{\rm DMAP}$, see  Eqs.~\eqref{eq:lnS} and \eqref{eq:PG}. We see from the figure, that these quantities are numerically very similar for the two tests, as well as for the two parameterizations. For the parameter goodness-of-fit test, the quantity $Q_{\rm DMAP}$ is obtained by taking $\hat\theta_D$ in Eq.~\eqref{eq:PG} as parameter value at the maximum of the posterior. For the $\Sigma$ parameterization, we have only one relevant parameter (namely $\sum m_\nu$) for which we take a flat linear prior. Hence, in this case maximum posterior (MAP) and maximum likelihood (MLH) are identical and therefore $Q_{\rm DMAP} \equiv \chi^2_{\rm PG}$. For the $3\nu$ parameterization we adopt flat priors in the logarithm of the three neutrino masses, constrained by hyper-parameters (see section~\ref{subs:parameterizations}). Hence, here in principle, MAP and MLH are not identical. However, we see from Fig.~\ref{fig:lnS-PG} and Tab.~\ref{tab:summary} that the $Q_{\rm DMAP}$ values for the $3\nu$ and $\Sigma$ parameterizations are very close, and hence we find also for $3\nu$ that the relationship $Q_{\rm DMAP} \approx \chi^2_{\rm PG}$ holds to good accuracy.

Under certain regularity conditions the quantities shown in the upper panels of Fig.~\ref{fig:lnS-PG} follow a $\chi^2_d$ distribution, with $d$ corresponding to the effective number of parameters in common to the two data sets, as defined in Eqs.~\eqref{eq:d} and \eqref{eq:n}, respectively. We give the Bayesian model dimensionalities obtained for the various data set combinations in the right part of Tab.~\ref{tab:summary}. We observe that in many cases Eq.~\eqref{eq:d} leads to negative values for $d$, which do not correspond to a meaningful $\chi^2$ definition. As we discuss in the Appendix, this follows from the properties of Bayesian model dimensionality and the specific shape of the posteriors in our application. Therefore, using Bayesian dimensionalities appears not suitable in our case to evaluate the effective number of degrees of freedom. Instead, we are going to use the simple parameter counting from Eq.~\eqref{eq:n} also in the case of the suspiciousness test, which gives $n=3$ or 1 for the $3\nu$ or $\Sigma$ parameterization, corresponding either to the 3 neutrino masses or to the singe parameter $\sum m_\nu$, respectively.\footnote{Using a $\chi^2_d$ distribution for $(d-2\ln S)$ in any case requires regularity conditions, such as Gaussian-shaped posteriors. Large deviations from $d_D$ from naive parameter counting signals non-Gaussian posteriors. The probabilities reported in Tab.~\ref{tab:summary} and lower panels of Fig.~\ref{fig:lnS-PG} have to be interpreted with care, and are understood {\it under the assumption} that $(d-2\ln S)$ follows a $\chi^2_n$ distribution with $n$ given in Eq.~\eqref{eq:n}.} 
We observe from the lower panels of Fig.~\ref{fig:lnS-PG} that, although the test statistics themselves are very similar, the tension quantified by the corresponding significance is somewhat stronger in the $\Sigma$ parameterization, due to the smaller number of dof. This effect is a known property of the parameter goodness-of-fit test: introducing more model parameters reduces the tension \cite{Maltoni:2003cu}.

Let us now discuss the physics results. Notice that current cosmological data (blue symbols) show mild tension with terrestrial data for NO at level of $1-2\sigma$ and for IO at the level of $2-3\sigma$. We can neither claim significant tension, nor disfavour IO due to strong tension.
Assuming a future determination of $\sum m_\nu$ of 0.06~eV according to the minimal NO value (green symbols), both tests show full consistency of the the data sets for NO and disfavour IO at the level of $2\sigma$. This is expected and in agreement with the trivial observation that for a determination according to Eq.~\eqref{eq:futureNO}, $\sum m_\nu \ge 0.1$~eV can be excluded at $2\sigma$. Note also, that the tension for IO in this case even slightly decreases with respect to current data, due to the finite mean value for $\sum m_\nu$. In particular, the PG test in the $3\nu$ parameterization signals a tension for IO only at the $1\sigma$ level, i.e., no tension. Moving now to the hypothetical case of no-mass detection of future cosmological data (magenta symbols), we see very strong tension for IO (above $4\sigma$), however, also significant tension for NO (between 2 and $3\sigma$). Hence, rejection of IO on the basis of this tension becomes problematic, as also the alternative hypothesis suffers from a non-negligible tension.

\subsection{Parameter shift test}

\begin{figure}
\centering
\includegraphics[width=0.7\columnwidth]{legend}\\
\includegraphics[width=0.99\columnwidth]{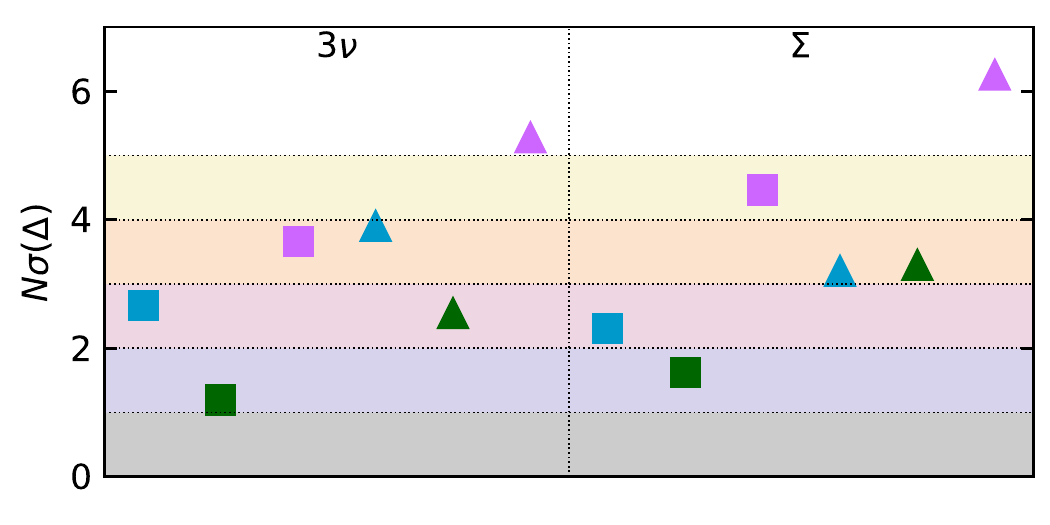}
\caption{Tension between cosmological and terrestrial experiments in terms of Gaussian standard deviations accordingly to the parameter shift test as defined in Eq.~\eqref{eq:delta_par_shift}. The left (right) part of the figure corresponds to the $3\nu$ ($\Sigma$) parameterization, see Sec.~\ref{subs:parameterizations}. Different colors indicate different assumptions on the  the cosmological data set, see Sec.~\ref{subs:data}, and square (triangle) symbols correspond to NO (IO) mass ordering.}
\label{fig:shifts}
\end{figure}

Figure~\ref{fig:shifts} depicts the corresponding results for the parameter shift test. In general we observe a similar pattern as for the suspiciousness and parameter goodness-of-fit tests, and the physics interpretation is similar. However, we notice in all cases that the parameter shift leads to a higher tension. According to the parameter shift test, current data shows tension of $\gtrsim 2\sigma$ ($\gtrsim 3\sigma$) for NO (IO). The non-observation of neutrino mass by future cosmology will lead to a (very) strong tension with oscillation data regardless of the mass ordering. Even for future NO, some tension close to the 2$\sigma$ level still remains for NO in the case of the $\Sigma$ parameterization. 

The reason for the relative stronger tensions obtained with the parameter shift test is a Bayesian volume effect. This test, as defined in Eq.~\eqref{eq:delta_par_shift}, measures the relative size of the overlap of the posterior volumes in parameter space of the two models. As an example, we can see from Fig.~\ref{fig:S_mnu_post} that even for the future NO case, the overlap volume with the terrestrial posterior is rather small. The result of the parameter shift test depends on the available parameter volume of the data sets, in particular on the upper bound on $\sum m_\nu$ from KATRIN: the tension will become stronger (weaker) for a weaker (stronger) upper bound on $\sum m_\nu$, just by increasing (decreasing) the terrestrial posterior volume in the region far away from the cosmological posterior volume.\footnote{We have checked that this effect is still relatively small if we use the final KATRIN sensitivity instead of the present result, for which results of the parameter shift test are rather similar.} Note also that there is no systematic trend when switching from the $3\nu$ to the $\Sigma$ parameterizations: while for current data the tension becomes weaker, for future NO as well as future 0 it becomes stronger (both for NO and IO).

\subsection{Mass ordering comparison}

Let us now briefly compare the tension measures presented above to a direct model comparison of NO versus IO. To this aim we consider the so-called Bayes factor, in analogy to the Bayesian evidence ratio from Eq.~\eqref{eq:lnZratio}:
\begin{equation}\label{eq:B}
B_{\rm NO,IO}\equiv
\frac{\mathcal{Z}_{\rm NO}}{\mathcal{Z}_{\rm IO}} \,.
\end{equation}
This quantity describes the Bayesian odds in favour of NO, i.e., large values of $B_{\rm NO,IO}$ correspond to a preference for NO. We convert Bayes factors into probabilities by using $P_{\rm NO}=B_{\rm NO,IO}/(1+B_{\rm NO,IO})$ and $P_{\rm IO}=1/(1+B_{\rm NO,IO})$ (given equal initial prior probabilities).

\begin{figure}[t]
\centering
\includegraphics[width=\columnwidth]{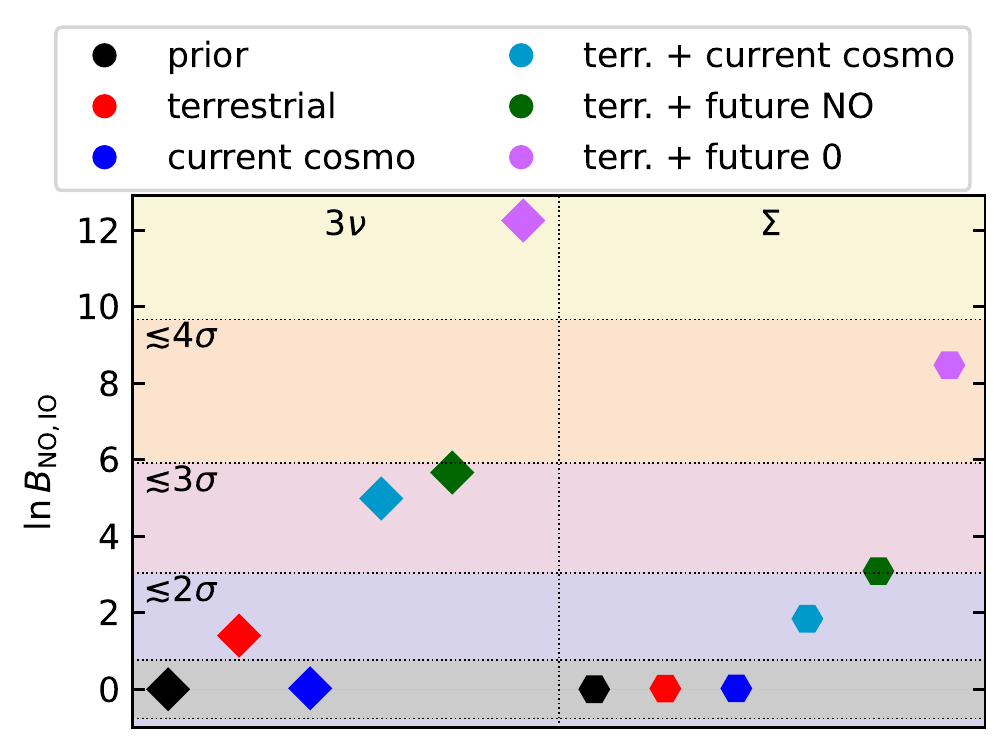}
\caption{Bayes factors in favor of NO for various data combinations and the two 
considered parameterizations. Significances indicated by the background shading correspond to the probability for IO obtained as $P_{\rm IO}=1/(1+B_{\rm NO,IO})$ converted into Gaussian standard deviations.}
\label{fig:lnBnoio}
\end{figure}

Figure~\ref{fig:lnBnoio} shows the logarithm of the Bayes factor. Here we show only the contribution from the available parameter space volume from the interplay of cosmological and terrestrial data, in order to compare with the tension measures discussed above. We note that here the direct contribution from the $\chi^2$ difference between NO and IO from oscillation data alone \cite{deSalas:2020pgw,Esteban:2020cvm,Capozzi:2021fjo} is not considered, which may provide additional NO/IO discrimination in the Bayes factor, see \cite{Gariazzo:2022ahe} for a recent discussion.

The black, red, and dark-blue symbols in the figure correspond to using the prior-only, terrestrial data alone, and current cosmology without terrestrial data, respectively. None of these cases shows any significant MO preference. Note that the slightly non-zero value for $\ln B_{\rm NO,IO}$ for terrestrial data in the $3\nu$ parameterization is a pure volume effect. By comparing the 
$\ln B_{\rm NO,IO}$ results for the combination of terrestrial and cosmological data (light-blue, green and magenta), we observe a significant dependence on the parameterization. This is in line with the arguments discussed in \cite{Gariazzo:2022ahe,Schwetz:2017fey}, where it is stressed that parameterizations with three independent neutrino masses in general lead to a strong preference for NO compared to other parameterizations. Indeed, from Fig.~\ref{fig:lnBnoio} we see approximately a difference of $1\sigma$ between the two considered parameterizations.

Concerning future cosmological data,
a measurement $\mnu=0.06\pm0.02$~eV would provide a significance
of approximately $2-3\sigma$ in favor of NO. Hence, from this argument alone (i.e., without using additional information from oscillation data), a precision such as the one considered here is not sufficient for a decisive determination of the mass ordering.
Within the case ``future 0'', for which the measurements provide a preferred value $\mnu=0$~eV,
the preference for NO is strong, close to the $4\sigma$ level (even for the $\Sigma$ parameterization).
This result, however, is a consequence of the stronger rejection of the region at $\mnu>0.1$~eV
with respect to the one at $\mnu>0.06$~eV, and does not take into account that also the NO solution suffers from a tension between cosmology and oscillation data,
as discussed in the previous sections.

\section{Conclusions}
\label{sec:conclusions}

The neutrino mass sensitivity from cosmological data analyses  is entering an exciting phase, approaching the minimal values for $\mnu$ as required by oscillation data, i.e., $\mnu \approx 0.06$ (0.1)~eV for NO (IO). 
In this manuscript we discuss quantitative measures to evaluate a possible tension between cosmology and terrestrial neutrino mass determinations. In particular we have applied the Bayesian suspiciousness test, parameter goodness-of-fit tests and Bayesian parameter differences, and studied implications for current cosmological data or sensitivities to be expected in the near future. In the latter case we assume an accuracy of 0.02~eV on $\mnu$ and consider two possible scenarios, either a mean value of $\mnu = 0.06$~eV, i.e. the minimal value predicted for NO, or $\mnu = 0$, i.e., a hypothetical non-observation of neutrino mass in cosmology. Our main conclusions can be summarized as follows:
\begin{itemize}
    \item Current data show modest tension between cosmology and terrestrial data, at the level of $1-2\sigma$ for NO and $2-3\sigma$ for IO. 
    \item If future cosmology will find a value of $\mnu\approx 0.06$~eV 
    (corresponding to NO with vanishing lightest neutrino mass) the tension for IO will be at the level of $2-3\sigma$. Hence, the assumed accuracy on $\mnu$ of 0.02~eV is not sufficient to exclude IO with decisive, strong significance.
    \item If future cosmological measurements do not find evidence for a non-zero neutrino mass, the tension with terrestrial data will be at the level of $2-3\sigma$ for NO and $\gtrsim 4\sigma$ for IO. Only in this case IO can be disfavoured strongly, however, at the price of having a tension between cosmology and terrestrial data present also for NO.  
    \item Bayesian suspiciousness and parameter goodness-of-fit tests give very similar results. In both cases, tension quantified in terms of significances depends on the number of model parameters. 
    \item We find that Bayesian model dimensionality is not a useful measure for the relevant degrees of freedom in our case of interest; our results are based on ``naive'' parameter counting.  
    \item Parameter differences in general show stronger tensions and depend on priors and parameterizations in a non-trivial way. 
\end{itemize}

In conclusion, in this work we have emphasized the well-known fact that the neutrino mass-ordering sensitivity in cosmological data analyses emerges from the available parameter space in the interplay between cosmology and neutrino oscillation data. In such a situation, the relative comparison of NO and IO in terms of model-comparison may be misleading. Excluding one of the hypotheses makes sense only if the alternative hypothesis provides a good fit to the data. We have used statistical tests as tension diagnostics between data sets in order to address this point. We argue that IO can only be excluded in a meaningful way if cosmology finds a result for $\mnu$ consistent with the NO prediction. Quantitatively we find that, based on this argument, an accuracy better than 0.02~eV from cosmological observations will be absolutely required in order to reject the inverted mass ordering with decisive significance.

If the tension between cosmological measurements and oscillation results is found also in NO case, or, eventually, a (positive) neutrino mass signal is detected in  either beta-decay or neutrinoless double beta decay experiments, but cosmological observations prefer $\sum m_\nu=0$, a modification of the neutrino sector or of the dark sector of the theory would be required (see also the discussion in Ref.~\cite{Gerbino:2022nvz}). In such a case, dark energy could have a phantom nature~\cite{Vagnozzi:2018jhn}, or the dark matter sector could show unexpected interactions, hiding the effect of neutrino masses~\cite{Stadler:2018dsa}.
On the other hand, a modification in the neutrino sector could also explain such potential tensions. Some examples are neutrino decay \cite{FrancoAbellan:2021hdb,Escudero:2020ped} or conversion  \cite{Farzan:2015pca,Escudero:2022gez} 
into dark radiation, long-range neutrino interactions~\cite{Esteban:2021ozz} or time-dependent neutrino masses \cite{Dvali:2016uhn}. 
It is therefore crucial to confront future cosmological mass limits with neutrino oscillation results, as a tool to constrain beyond the standard model interactions in the \emph{invisible} sector (neutrinos, dark matter and dark energy) of the theory.

\begin{acknowledgments}

This project has received support from the European Union’s Horizon 2020 research and innovation programme under the Marie Sklodowska-Curie grant agreements  
No 754496 (FELLINI) and No 860881 (HIDDeN). OM is supported by the MCIN/AEI/10.13039/501100011033 of Spain under grant PID2020-113644GB-I00, by the Generalitat Valenciana of Spain under the grant PROMETEO/2019/083 and by the European Union’s Framework Programme for Research and Innovation Horizon 2020 (2014–2020) under grant H2020-MSCA-ITN-2019/860881-HIDDeN.
\end{acknowledgments}

\appendix

\section{Properties of Bayesian model dimensionality}
For the calculation of the number of degrees of freedom in a Bayesian context,
the authors of \cite{Handley:2019pqx} suggest to employ the Bayesian model dimensionality $d_D$ defined in our Eq.~\eqref{eq:bmd}.
In their article, the authors show that a $1$-dimensional Gaussian distribution corresponds to $d_D=1$,
while different probability distributions may correspond to $d_D$ significantly different from 1, see their figure~3. For instance, a top-hat distribution gives $d_D=0$. We provide further examples below.
In our specific case, 
as we can see from the last four columns in table~\ref{tab:summary}, many of the $d_D$ values are quite significantly different from the naive parameter counting.

Let us first analyse the column $d_A$, which corresponds to terrestrial experiments.
In the $3\nu$ parameterization, we have $d_A\simeq2$, as expected from the fact that terrestrial experiments constrain two mass splittings, while the absolute scale of neutrino masses is not significantly constrained.
This is particularly clear from $d_A\simeq0.45$ in the $\Sigma$ case, for which the \mnu\ posterior partially resembles a 
top-hat distribution (see the red curves in figure~\ref{fig:S_mnu_post}),
that would correspond to $d_D=0$ according to \cite{Handley:2019pqx}.

The columns $d_B$ and $d_{AB}$ are more complicated to interpret,
but we can understand the results in this way.
Let us consider a one-dimensional Gaussian distribution, which gives $d_D=1$ for a sufficiently wide prior range including the maximum.
For example, if we consider a one-dimensional Gaussian on some parameter $x$, with $\mu=0$ and $\sigma=1$, we would still get $d_D=1$ if we consider a prior that cuts in half the distribution (such as, for example $0\le x\le 10$).
On the contrary, we can obtain $d_D<1$ if we consider an asymmetric range that includes $\mu$ (for example $d_D=0.67$ for $-1\le x\le 10$) or $d_D>1$ if the central value falls outside the interval (e.g.~$d_D=1.36$ for $1\le x\le10$).
Notice that the former case mimics the KATRIN likelihood from Eq.~\eqref{eq:mbeta}, since the central value falls inside the allowed prior range and we get $d_D<1$,
while the latter corresponds to the combination of terrestrial and cosmological data, for which the central value from cosmology ($\mnu=0$) is outside the \mnu\ range allowed by oscillation experiments ($d_{AB}>1$ in most of the cases).
The only case $d_{AB}=1$ is obtained when the preferred cosmological value is at the prior edge (``future NO'' scenario, fitted with NO).

We also noticed that switching from a linear to a logarithmic prior on the considered parameter(s) can generate significant differences in the value of $d_D$:
it is therefore reasonable to have much wider fluctuations in the Bayesian model dimensionality in the $3\nu$ scenario, which considers three log-normal distributions for the mass parameters, than in the $\Sigma$ case, for which we have a simple linear prior and one varying parameter.
Similar conclusions can be drawn for the $d_B$ column.

\bibliography{tensions}

\end{document}